\documentclass[conference]{IEEEtran}
\IEEEoverridecommandlockouts
\usepackage{cite}
\usepackage{amsmath,amssymb,amsfonts}
\usepackage{algorithmic}
\usepackage{graphicx}
\usepackage{textcomp}
\usepackage{xcolor}
\usepackage[pangram]{blindtext}

\def\BibTeX{{\rm B\kern-.05em{\sc i\kern-.025em b}\kern-.08em
    T\kern-.1667em\lower.7ex\hbox{E}\kern-.125emX}}

\usepackage[utf8x]{inputenc}

\makeatletter
\def\endthebibliography{%
  \def\@noitemerr{\@latex@warning{Empty `thebibliography' environment}}%
  \endlist
}
 \let\old@ps@headings\ps@headings
 \let\old@ps@IEEEtitlepagestyle\ps@IEEEtitlepagestyle
 \def\confheader#1{%
 \def\ps@headings{%
 \old@ps@headings%
 \def\@oddhead{\strut\hfill#1\hfill\strut}%
 \def\@evenhead{\strut\hfill#1\hfill\strut}%
 }%
 \def\ps@IEEEtitlepagestyle{%
 \old@ps@IEEEtitlepagestyle%
 \def\@oddhead{\strut\hfill#1\hfill\strut}%
 \def\@evenhead{\strut\hfill#1\hfill\strut}%
 }%
 \ps@headings%
 }
 \makeatother
 
\confheader{Preprint submitted to the 35\textsuperscript{th} International Conference on Software Maintenance and Evolution (ICSME'19)}

\begin{document}

\title{Business-Driven Technical Debt Prioritization}

%----AUTHORS---------------

\author{\IEEEauthorblockN{Rodrigo Rebouças de Almeida}
\IEEEauthorblockA{\textit{Federal University of Rio Grande do Norte (UFRN) - Natal, Brazil} \\
\textit{Federal University of Paraíba (UFPB) - Rio Tinto, Brazil} \\
rodrigor@dcx.ufpb.br}
	\thanks{Supervisor: Dr. Uirá Kulesza (uira@dimap.ufrn.br) - Federal University of Rio Grande do Norte (UFRN), Natal,  Brazil. Co-supervisor: Dr. Christoph Treude (christoph.treude@adelaide.edu.au) - University of Adelaide, Adelaide, Australia}

}

\maketitle

\begin{abstract}
Technical debt happens when teams take shortcuts on software development to gain short-term benefits at the cost of making future changes more expensive. Previous results show that there is a misalignment between the prioritization done by technical professionals and the prioritization expected by business ones. This paper presents a business-driven approach to prioritize technical debt items. The research is organized into four phases: exploratory, to identify the research focus; concept verification, where the proposed approach was evaluated on a multi-case study; solution, where a design science research was conducted to develop Tracy, a framework for technical debt prioritization; and validation. Results so far show that the business-driven prioritization of technical debt items can improve the alignment and communication between the technical and business stakeholders.
\end{abstract}

\section{Introduction}

Technical debt is a problem in software development and evolution that occurs when teams take a shortcut to gain short-term benefits at the cost of making future changes more expensive or impossible~\cite{Kruchtenbook2019}. Management and business factors are the leading causes of technical debt, and many papers have pointed out that research in this area should give more focus to the business aspects of technical debt~\cite{AMPATZOGLOU:2015,rios2019,terese2019}.

In this paper, we present a business-driven approach to improve technical debt prioritization. The research is being conducted in four stages: 1) \textit{exploratory} - research activities to identify and clarify the research problem; 2) \textit{concept verification} - where we proposed a business-driven approach to prioritize technical debt and verified if it would improve the technical debt prioritization; 3) \textit{solution} - where we developed \textit{Tracy} - a business-driven technical debt prioritization framework that supports our approach; and, 4) \textit{validation} - where we run \textit{Tracy's} preliminary validation.

The results so far show that: (i) the proposed business-driven approach to prioritize technical debt can be useful to support the prioritization of technical debt~\cite{icsme2018}; (ii) that the \textit{Tracy Framework} allows technical and business stakeholders to align their expectations of the technical debt prioritization, through a standard set of business metrics and a common view of the business processes~\cite{icsme2019}. Next steps include more effort on the validation phase by evolving and applying our results in more scenarios and companies.

To the best of our knowledge, this is the first research effort which proposes a technical debt prioritization approach that considers business processes and business metrics to support decision making. Business processes are what companies do to deliver value to customers. For example, a “sales” process in an e-commerce company is the set of activities, decisions, and events that must happen to allow the customer to buy products~\cite{bpmbookdumas}.

The proposed solution is a result of several interviews and surveys with practitioners, two case studies with two companies~\cite{icsme2018} and the use of design science research~\cite{dsr.book} with three companies.

The objective of this research is to answer the following research questions:

\textbf{RQ1} Can the business perspective improve the technical debt prioritization decision making?

\textbf{RQ2} How can the business perspective help the technical debt prioritization decision making?

In order to answer RQ1 and RQ2, we conducted a multiple-case study to evaluate a business-driven approach to technical debt prioritization; and we conducted Design Science Research (DSR) to develop a solution for the following design goal/problem statement~\cite{dsr.book}:

Improve technical debt prioritization by designing a business-oriented decision-making framework to promote the alignment between technical decisions and business expectations.

\section{Methodology and results}
\label{section.methodology}

The overall research is divided into four phases: 1) exploratory, 2) concept verification, 3) solution, and 4) validation:

The objective of the \textbf{exploratory} phase was to identify the focus and clarify the research problem. It was composed of the following three activities. 

\textbf{1.1) Literature review:} the review of the state of the art to identify the open research problems related to technical debt management; 
As a result, we identified a lack of studies considering the business perspective to technical debt management decisions.

\textbf{1.2) Interview with practitioners:} as a second step, we conducted interviews and focus groups with professionals from 15 global software companies, from 13 different industries, to understand how the industry addresses the technical debt, its causes, and business impacts. Based on the interviews, we verified that the technical debt problem is universal, is recognized by all interviewed practitioners, and that business forces are relevant drivers to the creation of technical debt. We also identified numerous types of technical debt and various scenarios where technical debt creates business impact.

\textbf{1.3) Survey with practitioners:} after the interview analysis, we double-checked the findings with a global survey about technical debt management and its causes and effects on the business.

After the interviews with practitioners, we decided to focus the research on the business-driven prioritization of technical debt. In the next phase (\textbf{concept verification}), we verified if a business-driven prioritization would contribute to the prioritization of technical debt.

\textbf{2) Concept verification:}  The results from phase one led us to propose a technical debt management approach that evaluates business value through business processes. To evaluate the approach, we conducted a multiple-case study \cite{case.study.yin} in two companies. Results show that the proposed business-driven approach can improve the prioritization of technical debt, considering business expectations~\cite{icsme2018}.

After verifying that the business-driven approach could help the prioritization of technical debt, we moved to the \textbf{solution} phase, organized as design and validation steps.

\textbf{3) Solution:} to develop a solution for the business-driven prioritization problem, we conduct design science research \cite{dsr.book} with three companies;  We developed a framework for business-driven technical debt prioritization with the participation of 49 professionals in 12 different groups from three companies, during six months. At this moment, we finished the design phase of the proposed framework (but as natural in a DSR, the solution can be evolved during all of its life-cycle).

\textbf{4) Validation:} the next step towards the end of the research is the application of Technical Action Research \cite{dsr.book} to validate if the solution meets the design requirements. We executed an initial evaluation which shows that the developed framework is coherent in its structure, but more cases must be executed to make the evidence stronger~\cite{icsme2019}.

\section{Results so far}

In this section, we present a \textit{business-driven prioritization approach} and the \textit{Tracy Framework}, which are the results of the \textit{concept verification} and \textit{solution} phases of this research, respectively. 

\subsection{Result 1: Business-driven prioritization approach}
\label{section.1stapproach}

The business process management (BPM) involves the lifecycle of the business processes and a set of management tools to deal with short-term to long-term aspects of the business~\cite{bpmbookdumas}. The proposed approach~\cite{icsme2018} relates technical debt to business processes through the BPM discipline to bring business values to the prioritization of technical debt.

\begin{figure}
\begin{center}
\includegraphics[width=0.5\columnwidth]{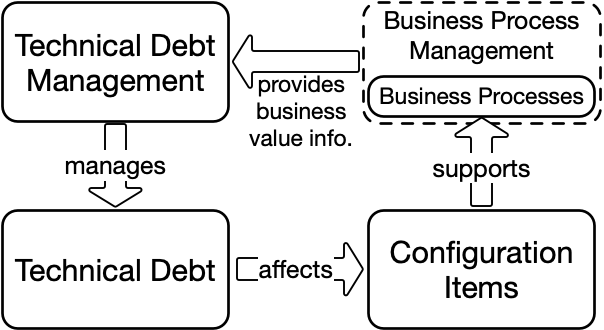}
\caption{Proposed approach for business-driven technical debt management.}
\vspace{-.5cm}
\label{fig_approach}
\end{center}
\end{figure}

Figure~\ref{fig_approach} shows the business-driven approach's main concepts and how they relate to each other. The \textit{technical debt management} uses information from \textit{business process management} to manage \textit{technical debt}, e.g., by prioritizing it. The technical debt affects \textit{configuration items} (managed IT artifacts, e.g., code, documentation, modules, or services). These configuration items support \textit{business processes}, from where it is possible to identify the business value and impact, which can finally close the cycle and support TD Management decision making.

\begin{figure}
\begin{center}
\includegraphics[width=0.6\columnwidth]{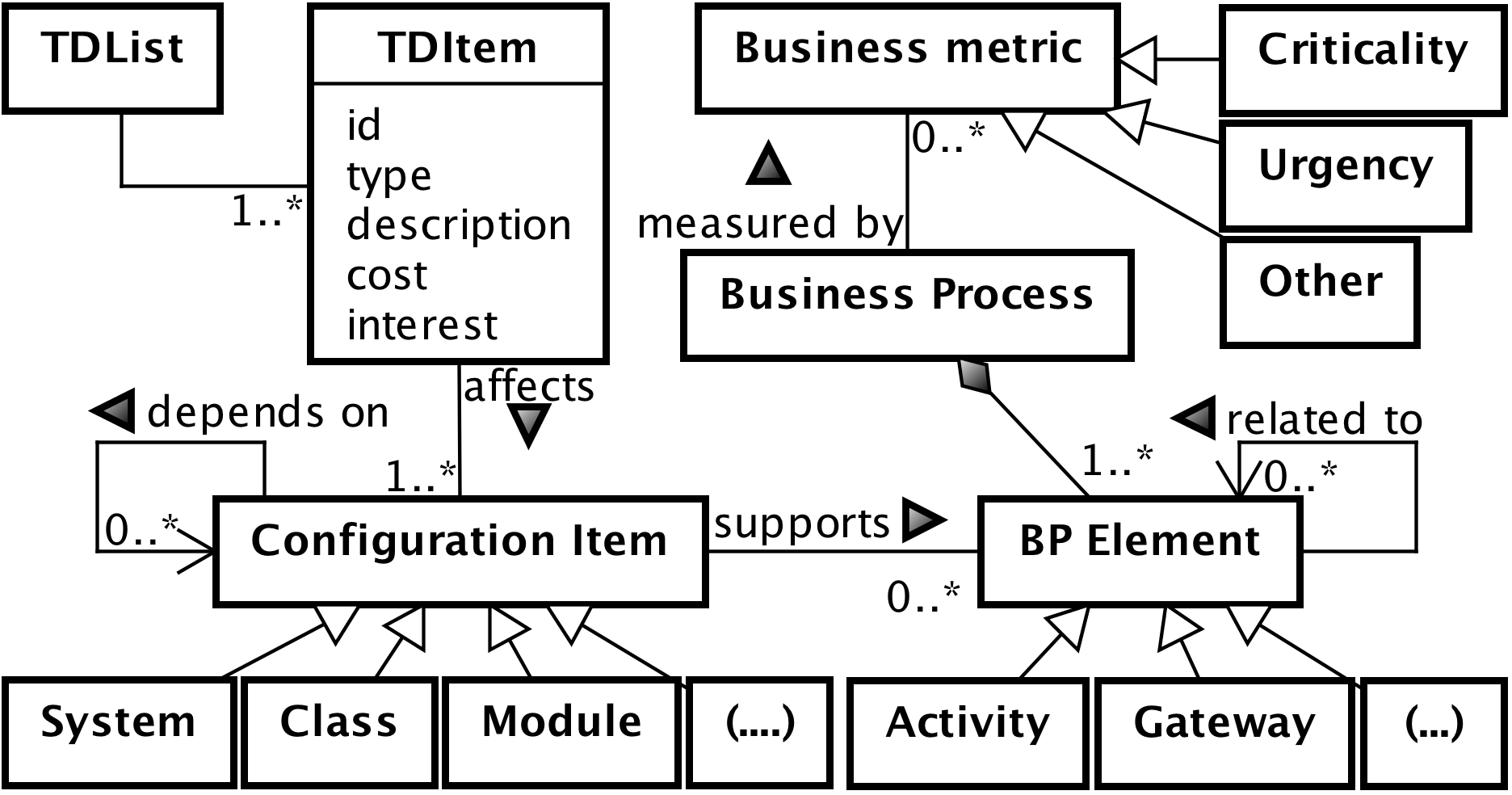}
\caption{Relationships between the technical debt list and the business process}
\vspace{-.5cm}
\label{fig_uml}
\end{center}
\end{figure}

Figure~\ref{fig_uml} shows our conceptual model of how technical debt items and business processes are related to each other. The model shows that a technical debt list (TDList) is related to one or more technical debt items (TDItems) which affect one or more (Configuration Items). A configuration item can support different business processes. A BP Element can have its priority and criticality evaluated in business terms (BP). BP Elements compose the (Business Process), which also has its overall priority and criticality. This model extends the conceptual model presented by Rios et al.~\cite{Rios2018} by adding the business process perspective.

\subsubsection{The prioritization}

To apply the approach to technical debt prioritization, it is necessary: (i) to keep track of a technical debt list; (ii) to relate debt items to configuration items; (iii) to identify the business processes which are supported by the configuration items; (iv) to identify which business metrics contribute to decision making (criticality, urgency, financial aspects, etc.); (v) for each business process: to classify it considering business metrics; and, finally, (vi) to conduct the technical debt prioritization using the business perspective.

In our first study we considered two high-medium-low business metrics which must be estimated by business stakeholders: (i) criticality - a perception of how critical a business process is for the company or its customers; and (ii) urgency - an evaluation about how urgently a problem with a business process must be solved.  

To conduct the prioritization, we mapped the identified business processes to the list of technical debt items using the configuration items, then we compared the technical debt items considering the business impact of the affected business processes.

\begin{table}
\begin{center}
\caption{Case study: Technical impact versus business impact}
\includegraphics[width=0.5\columnwidth]{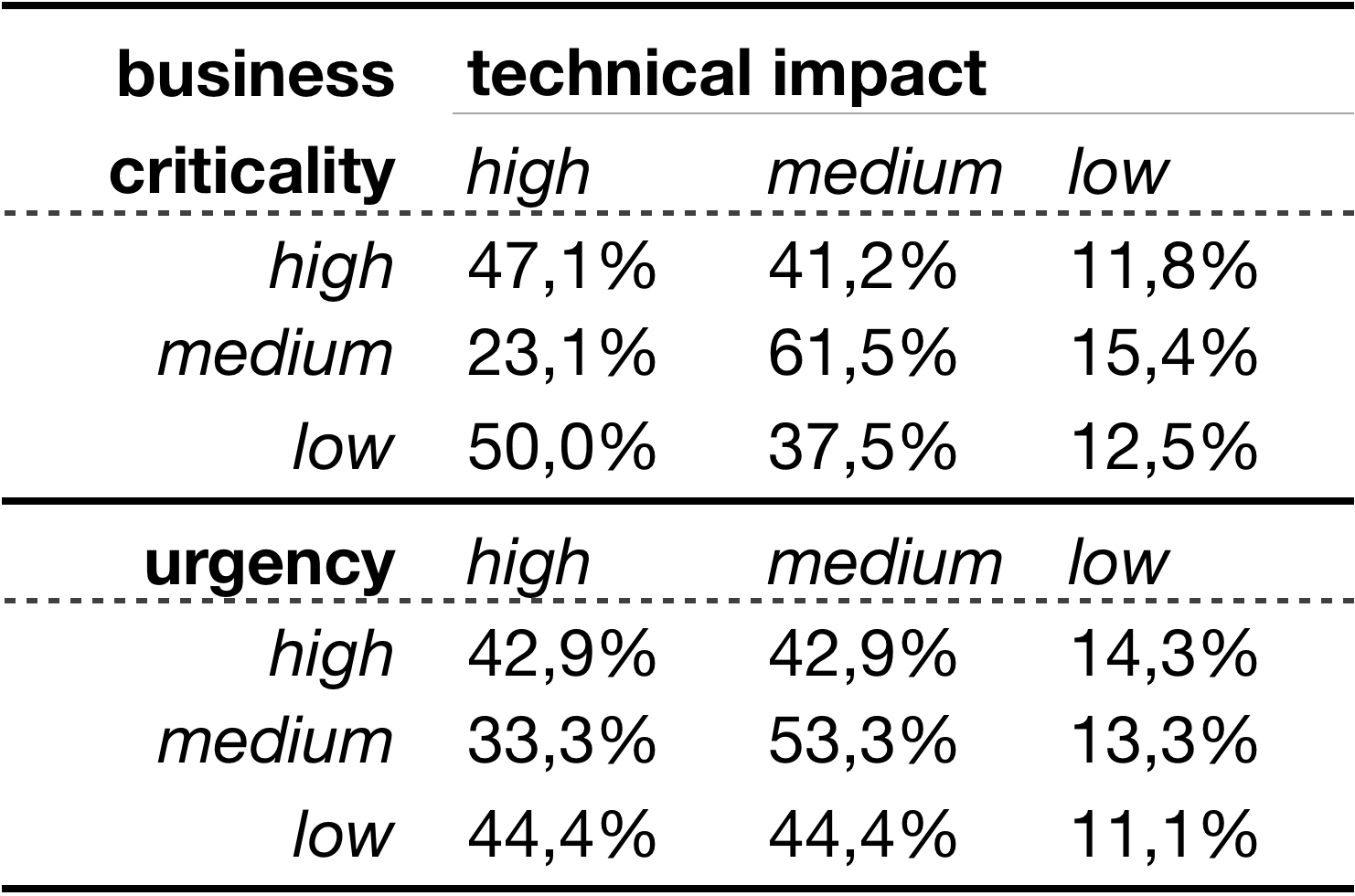}
\vspace{-.5cm}
\label{table_c1_impactVsBusiness}
\end{center}
\end{table}

To evaluate the result we used real data from two companies and compared a list of technical debt items prioritized considering a technical perspective with the same list prioritized using a new business perspective. Table~\ref{table_c1_impactVsBusiness} shows the percentage of technical debt item priorities which matched the business expectation, for one of the studied cases~\cite{icsme2018}. It shows how misaligned this decision would be with business objectives if the team prioritize the technical debt considering only a technical perspective. 

In the case study shown in Table~\ref{table_c1_impactVsBusiness}, regarding business criticality, 65\% of the technical debt items classified as high priority matched the business expectation. The same applies to 34.8\% of the medium priority items and 25\% of the low priority items. In total, the technical prioritization matched only 48.7\% of the criticality prioritization and only 35\% matched the urgency expectation.

This result provides evidence on how different a purely technical prioritization could turn out if it had been conducted from a business perspective, and that a business-driven prioritization, through the business process perspective, can be useful to support the prioritization of technical debt.

\subsection{Result 2: Tracy Framework}

Using our previous approach~\cite{icsme2018}, we developed the Tracy Framework~\cite{icsme2019} - a business-driven framework that prioritizes technical debt. To prioritize technical debt, \textit{Tracy} considers how IT assets (configuration items, e.g., systems or services, that create business value) support company's business processes. It has two major benefits: (i) to encourage different stakeholders to consider and identify business metrics that support decision making about technical debt, and (ii) to provide a prioritization mechanism that can be applied to different business and development contexts.

The development of the proposed framework is at the beginning of the third phase of Design Science Research (DSR)~\cite{dsr.book}, which was divided into the three phases of exploration, engineering and evaluation. The exploration and engineering phases involved the participation of 49 professionals from 12 different groups of three companies.

\begin{figure}
\begin{center}
\includegraphics[width=.9\columnwidth]{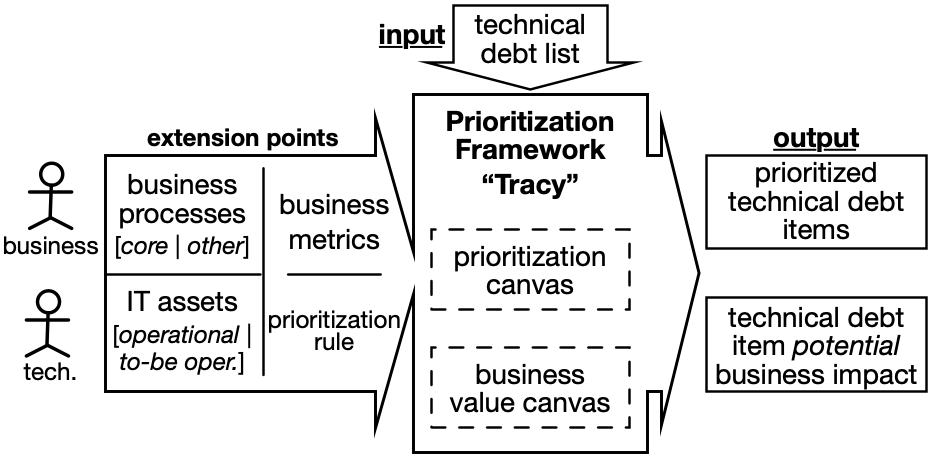}
\caption{Components of the technical debt prioritization framework}
\vspace{-.5cm}
\label{fig_tracy}
\end{center}
\end{figure}

The prioritization is done using the following information, shown in Fig.~\ref{fig_tracy}: (i) \textit{Business processes (BPs)} supported by IT assets - the processes are classified as \textit{core/support} or \textit{other}; (ii) a list of \textit{IT assets} that support the business processes; (iii) \textit{prioritization rules} to prioritize technical debt considering the impact of the IT assets on their supported business processes; and (iv) \textit{business metrics} related to each business process and IT asset.

\begin{figure}
\begin{center}
\includegraphics[width=.8\columnwidth]{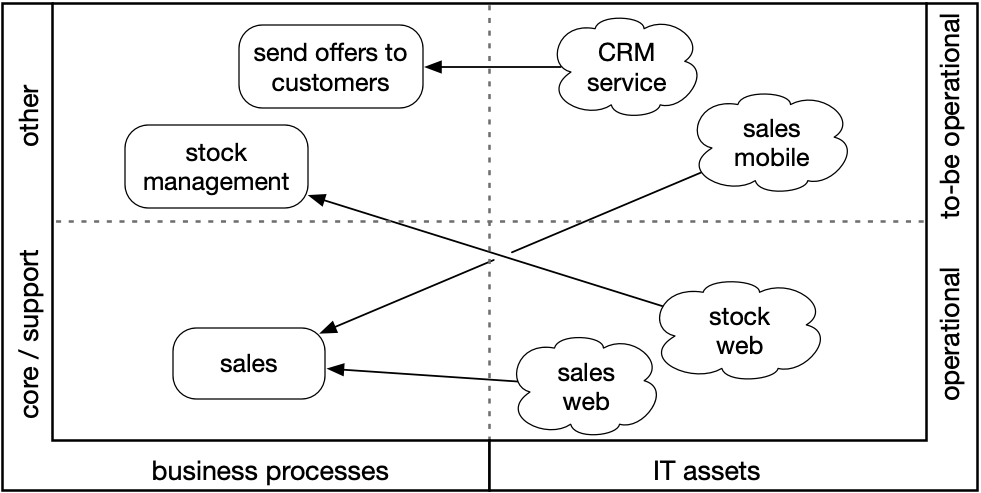}
\caption{Technical debt business-driven prioritization canvas}
\vspace{-.5cm}
\label{fig_prioritization_canvas}
\end{center}
\end{figure}

\textit{The Prioritization canvas} (Fig.~\ref{fig_prioritization_canvas}) is a board we have created to visualize the relationship between IT assets and their supported business processes. It is composed of four quadrants, where the business processes and their supported IT assets are arranged according to their types and states. On the left side are the business processes, categorized as \textit{core/support} or \textit{others}. On the right side are the IT assets, grouped into \textit{operational} and \textit{to-be operational}. The arrows express dependencies between them, e.g., the \textit{Sales} business process depends on the \textit{Sales web} system.

Different from the case study shown in Subsection \ref{section.1stapproach}, now the prioritization of technical debt items follows the relationship between IT assets and their supported business processes. Each company or project must decide what will guide the prioritization by defining a priority for each combination of business process type and IT asset operational state. Table~\ref{table_priority} shows an example of a prioritization rule where technical debt items that affect IT assets supporting \textit{core/support} business processes have a higher priority than others. Each company or even each project within the same company may have different prioritization rules.

\begin{table}
\begin{center}
\caption{Example of technical debt priority considering the IT assets and their supported business processes}
\includegraphics[width=.7\columnwidth]{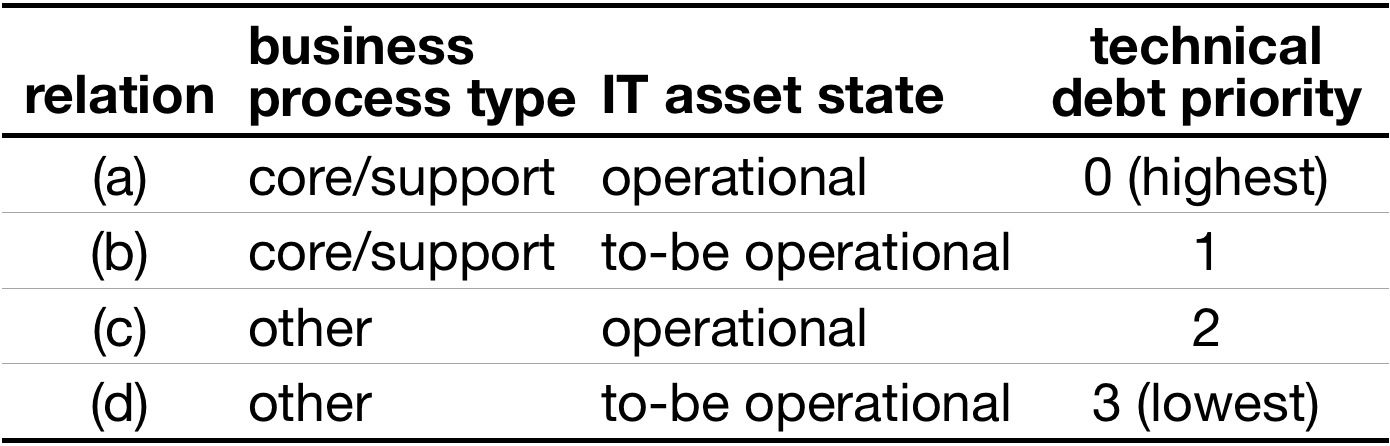}
\vspace{-.5cm}
\label{table_priority}
\end{center}
\end{table}

\begin{figure}
\begin{center}
\includegraphics[width=1\columnwidth]{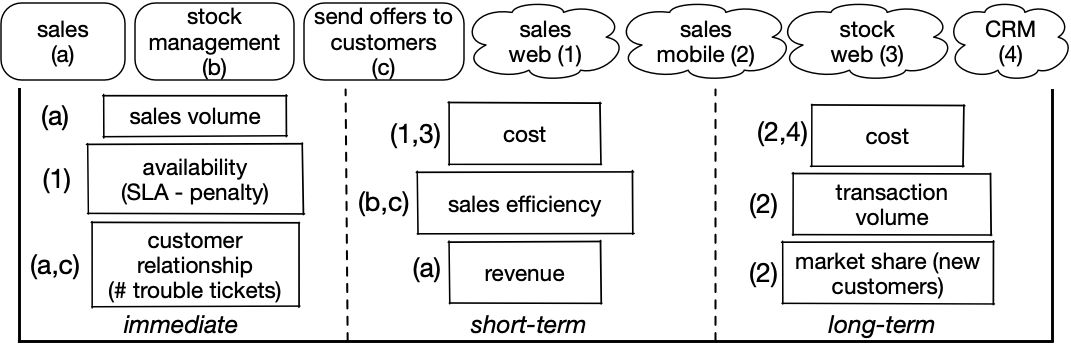}
\caption{Business-value canvas}
\vspace{-.5cm}
\label{fig_businessvalue_canvas}
\end{center}
\end{figure}

The \textit{Business-value canvas} (Fig.~\ref{fig_businessvalue_canvas}) is where each business process and IT asset is related to business metrics. Each metric may have immediate, short-term or long-term business impact. This canvas is a tool to help stakeholders to identify and classify the business value created by business processes and IT assets. The canvas aims at determining what is the potential immediate, short-term, and long-term business impact of technical debt which affects an IT asset. Depending on the company or project strategy, the time periods can be different from `immediate, short-term, and long-term'.

To identify the metrics, one must consider technical debt as a risk factor which may affect the business value~\cite{allman2012}, i.e., for each business process and IT asset, one must identify how they affect business, objectively. For example, in Fig~\ref{fig_businessvalue_canvas}, \textit{a technical debt which affects} the \textit{sales} BP has an immediate potential business impact on \textit{sales volume}, and on the \textit{customer Relationship}, and, finally, has impact on the \textit{revenue} (in short term). At the current stage, the framework does not consider the technical debt type or complexity for the prioritization. A simple ``code debt'' will be grouped with a complex ``architectural debt'' if both affect operational IT assets that support core business processes. The stakeholders must decide which item should be prioritized (among those which have a higher business impact).

\textit{Technical Debt Prioritization}: Table~\ref{table_example_priority} shows an example of a high priority technical debt item, based on the rule in Table~\ref{table_priority}, it has the highest priority. Its potential business impact is retrieved from the business-value canvas (Fig.~\ref{fig_businessvalue_canvas}). 

\begin{table}
\begin{center}
\caption{Example of a high priority technical debt item}
\includegraphics[width=1\columnwidth]{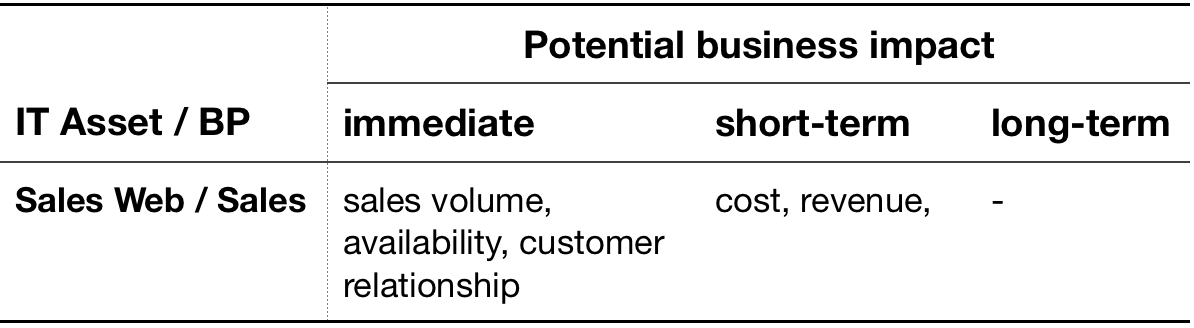}
\vspace{-.9cm}
\label{table_example_priority}
\end{center}
\end{table}

 The technical debt items with the highest priority should be the focus of decision making about ``which technical debt should we focus on now?'' Since the item in the example affects an operational service (\textit{Sales web}) that impact the \textit{Sales} business process, it should be investigated by verifying what type of technical debt it is and if it has immediate, short-term, or long-term potential business impact.

We can also see the potential business impact of technical debt items. For example, if one identifies a technical debt item that affects the sales volume or the system availability, then this item is a candidate to be paid immediately. If one identifies a technical debt item which affects cost (e.g., an inefficient algorithm that increases the cost of cloud infrastructure), the debt could be scheduled to be paid in the short-term. The third impact level (long-term) shows that debt should not ``sleep'' forever (e.g., an architectural issue that requires a considerable time and money investment to be paid).

\section{Related work and conclusions}

Several secondary and tertiary~\cite{AMPATZOGLOU:2015,Ribeiro:2016,MARTINI2018,FERNANDEZSANCHEZ201722,terese2019,RIOS2018-tertiary} studies analyze technical debt research. With regard to technical debt prioritization, it is a common finding that the criteria, tools, and approaches used to prioritize technical debt lack a business perspective. Lenarduzzi et al.~\cite{terese2019} conducted a systematic literature review on technical debt prioritization and identified only three papers that use business-related constraints. They highlight that based on most surveys conducted with practitioners, customer and business factors are the most important ones to consider when prioritizing technical debt. However, only a few papers addressed such factors. Besker et al.~\cite{besker2019} show that, among their surveyed and interviewed participants, there is no uniform way of prioritizing technical debt, and also that the prioritization process is highly dependent on individual practitioners’ influence. Our research contributes to normalizing the decisions about the prioritization, considering the business perspective. Ribeiro et al.~\cite{Ribeiro:2016} identified 14 decision-making criteria that can be used by development teams to prioritize the payment of technical debt items but only one of them considers the business aspect of cost-benefit analysis. 

We have presented Tracy - a decision-making framework that prioritizes technical debt considering how IT assets support a company's business processes, thus providing a new perspective on technical debt management. Information about the potential business impact of each technical debt item is crucial to support decisions among stakeholders with different roles. Tracy was constructed using Design Science Research~\cite{dsr.book}, with the participation of 49 practitioners over six months. 

Our future work includes more effort building and evaluating a method to guide the identification of business metrics which support the technical debt prioritization. We are currently developing a tool that partially automates the extraction of the information required by the framework (e.g., technical debt backlog and configuration items) by integrating with other development tools, like issue trackers, continuous integration tools and monitoring systems. This will enable us to apply our approach in many more scenarios and companies. The framework also opens up opportunities to advance research on the quantification of interest and principal of technical debt.

\textbf{Acknowledgements.} The author would like to thank his advisors, Dr. Uirá Kulesza, and Dr. Christoph Treude, for their invaluable guidance throughout his tenure as a doctoral student. He is also grateful to all of his collaborators from industry, particularly, 
André Figueiredo, 
Bruno Azevedo, 
Bruno Nóbrega, 
Carlos Ferreira,
Daniel Silveira,
Gregorio Melo,
Harley Barroso,
Isaura Lima, 
João Barbosa, 
José Cavalcanti,
Luiz Moura, 
Pedro Gomes,
Rauny Souza,
Rodrigo Breckenfeld, 
Rogério Viana,
Selene Lindsay, and
Vinicius Machado.

\bibliographystyle{IEEEtran}
\bibliography{bibliography/techdebt}

\end{document}